# Analysis of tunnel failure characteristics under multiple explosion loads based on persistent homology-based machine learning

Shengdong Zhang[1]  Shihui You*[1]  Longfei Chen[2]  Xiaofei Liu[2]

1. College of Mechanical and Electrical Engineering, Zaozhuang University 277160;

2. College of Civil Engineering and Mechanics, Xiangtan University 411105

**Abstract**

The study of tunnel failure characteristics under the load of external explosion source is an important problem in tunnel design and protection, in particular, it is of great significance to construct an intelligent topological feature description of the tunnel failure process. The failure characteristics of tunnels under explosive loading are described by using discrete element method and persistent homology-based machine learning. Firstly, the discrete element model of shallow buried tunnel was established in the discrete element software, and the explosive load was equivalent to a series of uniformly distributed loads acting on the surface by Saint-Venant principle, and the dynamic response of the tunnel under multiple explosive loads was obtained through iterative calculation. The topological characteristics of surrounding rock is studied by persistent homology-based machine learning. The geometric, physical and interunit characteristics of the tunnel subjected to explosive loading are extracted, and the nonlinear mapping relationship between the topological quantity of persistent homology, and the failure characteristics of the surrounding rock is established, and the results of the intelligent description of the failure characteristics of the tunnel are obtained. The research shows that the length of the longest Betty 1 bar code is closely related to the stability of the tunnel, which can be used for effective early warning of the tunnel failure, and an intelligent description of the tunnel failure process can be established to provide a new idea for tunnel engineering protection.

## 1 Introduction

In recent years, the field of tunnel engineering has developed rapidly, which has brought great convenience to people's travel. At the same time, the safety of tunnel construction and operation has also attracted more and more attention. Accidental explosions, terrorist attacks, etc. may cause more serious damage to the tunnel. At present, the most research which on the influence of explosion shock waves focus on the implosion sources,while the influence of external explosion sources on tunnels and the application of multiple explosive loads are relatively few. At the same time, because some targets with important strategic value are mostly hidden underground and protected by solid positive measures, a single missile strike is not effective. Therefore, in order to attack enemy underground targets, the United States and other countries have studied a large number of precision-guided weapons. This type of weapon has a characteristic that can allow multiple missiles to strike the same target, the first wave of strikes  blasts a gap firstly, followed by the first,  the second, third wave of strikes focused on killing. Therefore, studying the failure characteristics of tunnels under multiple loads of external explosion sources is of great significance for the design and evaluation of protection engineering, especially the development of tunnel safety early warning methods adapted to the effects of explosion loads [1].

There are also some difficulties to be solved in the study of tunnel failure characteristics. The main manifestation is that it is difficult to collect real-time data inside the rock mass structure in the experiment of the tunnel under the action of multiple and strong explosive loads, which affects the description and characterization of the degree of structural damage, and it is also difficult to analyze the topological characteristics with topology tools. Due to the complexity of the explosion process, the simulation is difficult to reach a perfect state. The above are the main difficulties restricting tunnel research. Therefore, the use of detection radar to directly detect the internal structure of the tunnel proposed in this paper, and then use the topology tool of continuous coherence to study the topological characteristics of the tunnel has

---

*<b>Corresponding author</b>

**E-mail address:** 101434@uzz.edu.cn

extremely important research significance and value.

The surrounding rock structure of the tunnel is often discontinuous, and its structure is cut by weak structural surfaces such as layers and joint surfaces. There are certain limitations in the study of continuum mechanics, while the discrete element method divides the surrounding rock into different units. The contact between the units can be separated. Taking into account the discontinuity of the joint surface of the tunnel surrounding rock, scholars mostly use the discrete element method to study the stability of the tunnel [2-6].

Persistent homology is a new branch of topological, it is based on algebraic topology and explores the topological characteristics of data by changing filter parameters (such as connected radius). Persistent homology builds a bridge between traditional geometry and topology, and shows great advantages in balancing data simplification and structural representation. In recent years, persistent homology has been continuously introduced in the fields of large data, artificial intelligence, intelligent manufacturing, material gene planning and deep learning [7-14]. Machine learning is the core of artificial intelligence, has been widely used in many fields with its maturing image processing and target recognition capabilities. However, machine learning has serious obstacles in characterizing high-dimensional and complex systems. Persistent homology-based machine learning not only solves the problem of machine learning dimensionality and complexity, but also retains the inherent topological characteristics of the data, and at the same time, the data can be processed intelligently. At present, the method of persistent homology-based machine learning has been successfully applied in the fields of dynamic and static image processing, medical imaging and diagnosis, protein ligands, classification and dynamics, and processing vibration time series analysis. Primoz[15] emphasized the combination of continuous coherence and machine learning for data analysis when summarizing topology tools.

However, the research of persistent homology-based machine learning in geotechnical engineering is rarely. Chen Longfei et al. [16] combined the particle discrete element and persistent homology theory to study the discretization of the slope of the accumulation body, and regarded the discretized block as a point cloud in the persistent homology generalized space, and obtained the persistent homology of the slope through the bar code graph. The relationship between the characteristics and the failure process proves the feasibility of persistent homology method to slope failure prediction.

This dissertation analyzes the displacement of the surrounding rock mass of the tunnel to study the failure characteristics of the surrounding rock of the tunnel. Regarding the surrounding rock of the tunnel as a multi-scale structure network, the bar code diagram of the surrounding rock structure of the tunnel is calculated by persistent homology, and the parameters of the one-dimensional bar code diagram are used to describe the failure characteristics of the surrounding rock of the tunnel. Finally, a support vector machine is introduced to predict the topological characteristics of the bar code map obtained by persistent homology, and obtain an intelligent description of the failure characteristics of the tunnel surrounding rock under the action of multiple explosions.

The specific implementation idea is to use ground penetrating radar to detect the displacement of rock masses around the tunnel, and use the obtained displacement data to calculate the persistent homology of the rock mass structure. Machine learning is carried out on the basis of persistent homology bar code graph, and support vector machine is introduced to intelligently describe the tunnel and predict the degree of damage.

# 2 The theory of persistent homology and improved support vector machine

## 2.1 Basic principles of persistent homology

Persistent homology is originated from Morsel theory, is a method used to calculate topological features at different spatial resolutions. Persistent homology can detect more continuous features on a wide spatial scale. These features are independent of the filter scale and can better represent the true characteristics of underlying space.

### 2.1.1 Simplicial Complexes

A simplicial complex is a combination of simplexes under certain rules. It can be viewed as a generalization of network or graph model.

Abstract Simplicial Complex

A simplex is the building block for simplicial complex. It can be viewed as a generalization of a triangle or tetrahedron to their higher dimensional counterparts.

**Definition 1** A geometric $k$-simplex $\sigma^k = \{v_0, v_1, v_2, \cdots, v_k\}$ is the convex hull formed by $k + 1$ *affinely independent points* $v_0, v_1, v_2, \cdots ; v_k$ in Euclidean space $R_d$ as follows, affinely independent points $v_0, v_1, v_2, \cdots ; v_k$ in Euclidean space $R_d$ as follows,

$$\sigma^k = \left\{ \lambda_0 v_0 + \lambda_1 v_1 + \cdots + \lambda_k v_k \left| \sum_{i=0}^{k} \lambda_i = 1; 0 \leq \lambda_i \leq 1, i = 0, 1, \cdots, k \right. \right\} \quad (1)$$

*A face $\tau$ of k-simplex $\sigma^k$ is the convex hull of a non-empty subset. We denote it as $\tau \leq \sigma^k$*

Geometrically, a 0-simplex is a vertex, a 1-simplex is an edge, a 2-simplex is a triangle, and a 3-simplex represents a tetrahedron. An oriented $k$-simplex $[\sigma^k]$ is a simplex together with an orientation, i.e., ordering of its vertex set. Simplices are the building block for (geometric) simplicial complex.

**Definition 2** *A geometric simplicial complex K is a finite set of geometric simplices that satisfy two essential conditions:*

*1. Any face of a simplex from K is also in K.*

*2. The intersection of any two simplices in K is either empty or shares faces.*

The dimension of $K$ is the maximal dimension of its simplexes. A geometric simplicial complex $K$ is combinatorial set, not a topological space. However, all the points of $Rd$ that lie in the simplex of $K$ aggregate together to topologize it into a subspace of $Rd$, known as polyhedron of $K$.

Graphs and networks, which are comprised of only vertices and edges, can be viewed as a simplicial complex with only 0-simplex and 1-simplex.

**Definition 3** *An abstract simplicial complex K is a finite set of elements $v_0, v_1, v_2, \cdots ; v_n$ called abstract vertices, together with a collection of subsets ($v_{i0}, v_{i1}, \cdots ; v_{im}$) called abstract simplexes, with the property that any subset of a simplex is still a simplex.*

For an abstract simplicial complex $K$, there exists a geometric simplicial complex $K0$ whose vertices are in one-to-one correspondence with the vertices of $K$ and a subset of vertices being a simplex of $K0$ if and only if they correspond to the vertices of some simplex of $K$. The geometric simplicial complex $K0$ is called the geometric realization of $K$

### 2.1.2 Homology[17]

Simplicial complexes $\kappa$, $K$ is summation of $\sum_{i=1}^{N} c_i [\sigma_i^k]$

Where $[\sigma_i^k]$ is the simplicial $K$ of simplicial complexes $\kappa$.

$c_i \in Z_2$, $Z_2$ represents a domain modulo 2, All k chains on $\kappa$ form an Abelian group, called chain group, denoted as

$c_k(K)$

For the edge operator $\partial_k$ of $K$ simplex $\sigma^k$, there are:

$$\partial_k \sigma_k = \sum_{i=0}^{k} (-1)^i = [u_0, u_1, ..., \hat{u}_i, ..., u_k] \quad (2)$$

Where, $[u_0, u_1, ..., \hat{u}_i, ..., u_k]$ means the face obtained by deleting the $i$-th vertex in the simplex. Edge operator can derive edge homomorphism $\partial_k : C_k(K) \to c_{k-1}(K)$.

The boundary operator has an important property: the composition operator $\partial_{k-1}\partial_k$ is the zero.

$$\partial_{k-1}\partial_k(\sigma^k) = \sum_{j<i}(-1)^i(-1)^j[u_0,\ldots,\hat{u}_i,\ldots,\hat{u}_j,\ldots,u_k] + \partial_{k-1}\partial_k(\sigma^k) = \sum_{j>i}(-1)^i(-1)^{j-1}[u_0,\ldots,\hat{u}_j,\ldots,\hat{u}_i,\ldots,u_k] = 0 \quad (3)$$

The chain group sequence connected by the boundary operator constitutes a chain complex, as shown below:

$$c_n(K) \xrightarrow{\partial_n} c_{n-1}(K) \xrightarrow{\partial_{n-1}} \cdots \xrightarrow{\partial_1} c_0(K) \xrightarrow{\partial_0} 0 \quad (4)$$

$\partial_{k-1}\partial_k = 0 \equiv \operatorname{Im}\partial_{k-1} \subset \operatorname{Ker}\partial_k$, where Im is image, Ker is kernel. The element of $\operatorname{Ker}\partial_k$ is called the $k$-th cyclic group, denoted as $Z_k = \operatorname{Ker}\partial_k$. The element of $\operatorname{Im}\partial_{k+1}$ is called the $k$-th edge group, denoted as $B_K = \operatorname{Im}\partial_{K+1}$, the k-th homology group is defined as the quotient group of $Z_k$ and $B_k$ [17]

$$H_k = Z_k / B_k \quad (5)$$

The $k$-th Betty number of the simplicial complexes $\kappa$ is a rank of $H_k$.

$$\beta_k = rank(H_k) = rank(Z_k) - rank(B_k) \quad (6)$$

Betty number $\beta_k$ is a finite number, because of $rank(B_p) \le rank(Z_p) < \infty$. The Betty number calculated by the homology group is used to describe the corresponding homology space. In general, the Betty numbers 1, 2, and 3 are the number of connectors, the number of holes surrounded by lines, the number of holes surrounded by faces, and the higher Betty numbers can be deduced by analogy.

### 2.1.3 Filtration and persistence

A fltration of a simplicial complex κ is a nested sequence of subcomplexes of κ

$$\phi = K_0 \subseteq K_1 \subseteq \ldots \subseteq K_m = K \quad (7)$$

With a fltration of simplicial complex κ, topological attributes can be generated for each member in the sequence by deriving the homology group of each simplicial complex. The topological features that are long lasting through the fltration sequence are more likely to capture signifcant property of the object. Intuitively, non-boundary cycles that are not mapped into boundaries too fast along the fltration are considered to be possibly involved in major features or persistence. Equipped with a proper derivation of fltration and a wise choice of threshold to defne persistence, it is practicable to flter out topological noises and acquire attributes of interest. The p-persistent $k$th homology group of $K_i$ is defned as[17]

$$H_k^{i,p} = Z_k^i / (B_k^{i+p} \cap Z_k^i) \quad (8)$$

where $Z_K^i = Z_k(K_i)$ and $B_k^i = B_k K_i$. The consequent $p$-persistent $k$th Betti number is $\beta_k^{i,p} = rank(H_k^{i,p})$. A well chosen $p$ promises reasonable elimination of topological noise.

### 2.1.4 Bar code image

The bar code plots is used to reflect the topological characteristics of the complex filtration duration of the point cloud set in the process of increasing the connected radius.

It is a collection of finite intervals on the real axis R, which can generally be expressed as [a,b] or [a ,+∞], where a,b∈R. If we draw the Betty interval in a two-dimensional coordinate system, we get a visual description of persistent homology. The traditional output of continuous coherence is a "barcode" plot, as shown in Figure 1:

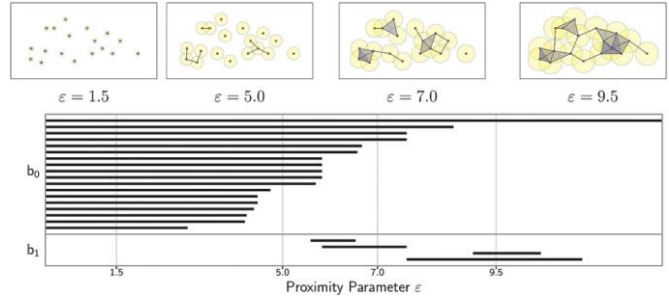

Fig1.Persist homology bar code

## 2.2 Least Squares Support Vector Machine Algorithm

Support vector machine is a machine learning method developed on the basis of statistical learning theory and has been widely used in geotechnical engineering. It has inherent advantages for solving small samples, high dimensionality, nonlinearity and local minima. However, traditional methods need to solve a quadratic programming problem, which is often very slow and computationally complex. The improved support vector machine transforms the inequality constraints in the support vector machine into equality constraints, ensuring the accuracy and speed of learning. The derivation of the improved support vector machine is as follows[18]:

$$\min_{w,b,e} F(w,b,e) = \frac{1}{2} w^T w + \frac{1}{2} \gamma \sum_{i=1}^{m} e_i^2 \quad (9)$$

Where, $e=[e_1,e_2,…,e_m]$ is deviation vector; $\gamma$ is weight; Equation (9) must meet the following constraints:

$$y_i = \left[w^T \varphi(x_i) + b\right] = 1 - e_i, i = 1,2,...,m \quad (10)$$

Define the Lagrange function and solve the maximum condition of the function, which is the minimum condition of equation (9). The Lagrange function is:

$$L(w,b,e,a) = F(w,b,e) - \sum_{i=1}^{m} \alpha_i \left[ y_i \left[ w^T \varphi(x_i) + b \right] - 1 + e_i \right] \quad (11)$$

Where, $\alpha_i$ is Lagrange multiplier, the optimal conditions are:

$$\begin{aligned}
\frac{\partial L}{\partial w} &= 0 \Rightarrow w = \sum_{i=1}^{m} \alpha_i y_i \varphi(x_i) \\
\frac{\partial L}{\partial b} &= 0 \Rightarrow \sum_{i=1}^{m} \alpha_i y_i \\
\frac{\partial L}{\partial e_i} &= 0 \Rightarrow \alpha_i = \gamma e_i, i = 1,2,...,m \\
\frac{\partial L}{\partial \alpha_i} &= 0 \Rightarrow y_i \left[ w^T \varphi(x_i + b) - 1 + e_i \right] = 0, i = 1,2,...,m
\end{aligned} \quad (12)$$

Equation (12) is transformed into the following linear equation

$$\begin{bmatrix} I & 0 & 0 & -Z^T \\ 0 & 0 & 0 & -Y^T \\ 0 & 0 & \gamma I & -I \\ Z & Y & I & 0 \end{bmatrix} \begin{bmatrix} w \\ b \\ e \\ \alpha \end{bmatrix} = \begin{bmatrix} 0 \\ 0 \\ 0 \\ I \end{bmatrix}$$

Where, $Z = \left[\varphi(x_1)^T, \varphi(x_2)^T y_2, \ldots, \varphi(x_m)^T y_m\right]$; $Y = [y_1, y_2,\ldots, y_m]$; $I = [1,\ldots,1]$; $e = [e_1, e_2,\ldots, e_m]$; $\alpha = [\alpha_1, \alpha_2,\ldots, \alpha_m]$。

According to the above derivation, it can be seen that the improved support vector machine converts the inequality constraints in the support vector machine into equality constraints, and the training process is also transformed into the solution of linear equations, which simplifies the complexity of calculations and ensures accuracy.

### 2.3 Selection and establishment of topological features

The tunnel is deformed by the explosive load. It is necessary to find its topological characteristics in the persistent homology, and search for useful features from these topological characteristics, so as to find the failure characteristics of the tunnel surrounding rock. Since we are building a two-dimensional model of the tunnel, only need to find the features corresponding to b1 and b2.

**Table 1: A list of features used in support vectors**

| Feature | Betti | Description |
| --- | --- | --- |
| 1 | 0 | The sum of the length of all 0-dimensional barcode images |
| 2 | 1 | The sum of the lengths of all 1-dimensional Betty numbers |
| 3 | 0 | The length of the second longest Betti 0 bar |
| 4 | 0 | The length of the third longest Betti 0 bar. |
| 5 | 0 | The summation of lengths of all Betti 0 bars except for those exceed the max fltration value. |
| 6 | 0 | The average length of Betti 0 bars except for those exceed the max fltration value. |
| 7 | 1 | The onset value of the longest Betti 1 bar. |
| 8 | 1 | The length of the longest Betti 1 bar. |
| 9 | 1 | The smallest onset value of the Betti 1 bar that is longer than 1.5 times |
| 10 | 1 | The average of the middle point values of all the Betti 1 bars that are longer than 1.5 times |
| 11 | 1 | The sum of the lengths of all 1-dimensional Betti numbers beyond the filter value |
| 12 | 1 | The average value of the barcode length of the 1-dimensional Betti number exceeding the filter value |
| 13 | 0 | Sum of the number of 0-dimensional bar code graphs |
| 14 | 1 | Sum of the number of bars in a 1-dimensional bar code graph |

**Table. 2 Categories corresponding to features**

| Feature | Feature number |
| --- | --- |
| Interaction strength and distribution between units | 1~2 |
| | 13~14 |
| Physical characteristics | 3~6 |
| Geometric Features | 7~12 |

Table 1 shows the 14 features commonly used in support vector machines. The second column indicates whether each feature corresponds to a 0-dimensional or a 1-dimensional Betty number. The last column is an explanation of the specific topological significance of this feature. Since the features in Table 1 are divided into 14 items, which are relatively detailed and complex, Table 2 classifies 14 features from a large level, and is divided into 3 categories, reflecting the intensity distribution, physical characteristics and geometric characteristics respectively.

## 3 Dynamic response of tunnel under explosive load

### 3.1 Discrete element explosion model

The discrete element model was established by referring to the actual situation of Chongqing Wulukou Metro Station [19]. After exploring, the surrounding rock of the tunnel is mostly composed of oil sandstone, which belongs to Grade IV. The influence of tunnel depth and boundary which effect on the deformation of tunnel surrounding rock is considered, the model as shown in Fig.2.

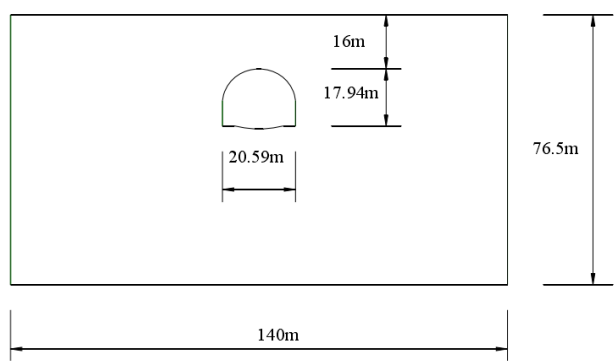

Fig.2 Model size

The width of the model is $l$=140m, the height is $h$=76.5m, the top buried depth of the excavated tunnel is $f$=16m, the width of the excavated tunnel section is $l_1$=20.59m, and the excavation height is $h_1$=17.49m. The mechanical parameters and joint mechanical parameters of the tunnel surrounding rock are selected in strict accordance with the material parameters corresponding to the grade of the surrounding rock in the engineering geological report, as shown in Table 3 and Table 4. In the simulation, the blasting action will cause the deformation of the rock mass, so the Mohr-Coulomb criterion is used when setting the rock mass and joints. According to literature [19], when the joint plane spacing is 5m, the failure area, plastic area and average displacement of the surrounding rock are the largest. Therefore, this research selects two joints that penetrate the tunnel surrounding rock with a joint spacing of 5m and a dip angle of 50°/100°. In order to make the model boundary not affect the accuracy of the simulation, after making the model reach an equilibrium state under the action of gravity, viscous boundary conditions are imposed on the bottom and both sides, horizontal constraints on both sides, and vertical constraints on the bottom.

**Tab.3 Mechanical parameters of tunnel surrounding rock**

| Density | bulk modulus | shear modulus | Angle of internal friction | Cohesive force | Tensile strength |
|---|---|---|---|---|---|
| $kg/m^3$ | GPa | GPa | ° | MPa | MPa |
| 2300 | 4.17 | 1.92 | 35 | 0.5 | 1.0 |

**Tab.4 Kinetic mechanics parameters**

| Normal stiffness | Tangential stiffness | Angle of internal friction | Cohesive force | Tensile strength |
|---|---|---|---|---|
| Gpa | Gpa | ° | Mpa | Mpa |
| 10 | 10 | 20 | 0.04 | 0 |

It is assumed that the external explosion source is directly above the tunnel to simulate the impact of the ground explosion on the unsupported tunnel. When the simulated tunnel is subjected to the explosive load from the ground, it is equivalent to applying a uniform load that varies with time on the top surface of the model [20]. In the process of numerical simulation, the explosion load time history curve is mainly processed into triangular, exponential and other forms [21]. This paper adopts the triangular form, and the duration of the explosion load is usually considered to be 1×10-6~0.1s [22]. The comprehensive consideration model is shallow buried compared with the calculation method of continuum mechanics in the tunnel and discrete element calculations, the stress wave is relatively slow in the internal transmission process, so the duration of the explosion load is 35ms, of which the pressure rise time is 5ms, the positive pressure action time is 30ms, at the same time, the value of the explosion load is relatively large to simulate the maximum damage , the derivation of the load is as follows.

$$P = \frac{2R}{L} \cdot P_0 \quad (13)$$

Where, $P$ is the equivalent uniform load, $R$ is the radius of the blast hole, and $L$ is the blast hole spacing. The blast hole spacing of the model in this paper is 1.25m, and $P_0$ is the load generated by a single blast hole.

$$P(t) = P_m \cdot f(t) \quad (14)$$

$$P_m = \frac{1}{8} \cdot \rho_c \cdot D^2 \cdot \kappa_c^{-6} \cdot \eta \quad (15)$$

Where, $f(t)$ is the exponential function of the explosion load with time, $P_m$ is the peak of the explosion load, $D$ is the explosive explosiveness, $\rho_c$ is the charge density, $K_c$ is the uncoupling coefficient, and $\eta$ is the enlargement factor which hole wall pressure affected by the detonation gas, $K_c$ = 9.

So, the formula for uniform load is

$$P = \frac{R}{4L} \cdot \rho_c \cdot D^2 \cdot \kappa_c^{-6} \cdot \eta \cdot f(t) \quad (16)$$

According to reference [20], $P_m$ and $p$ are calculated as

$p_m = 1.5 \times 10^9 Pa$ , $p = 1.38 \times 10^8 f(t)$ .

## 3.2 Dynamic response of the tunnel

When the local foundation soil and the tunnel structure are fixed, under the action of the explosion load, the maximum displacement of the tunnel structure increases with the increase of the explosion load intensity and the time of the explosion load, and decreases with the increase of the buried depth of the tunnel, showing an approximately linear relationship [23]. According to the situation of this article, it can be seen that the tunnel will have a large deformation due to the fact that the explosion load is large and repeated, the tunnel surrounding rock is not supported and reinforced, and the rock mass is sandstone. Therefore, we applied 20 explosion loads to the model, and the displacement change table is listed in the table 5.

**Table 5 Tunnel displacement table under multiple explosion loads**

| Explosion/time | Displacement | |
| --- | --- | --- |
| | Maximum displacement | Displacement of the upper collapse zone of the tunnel |
| 0 | 0.232 | 0.208 |
| 1 | 0.264 | 0.212 |
| 2 | 0.282 | 0.249 |
| 3 | 0.331 | 0.253 |
| 4 | 0.389 | 0.272 |
| 5 | 0.788 | 0.488 |
| 6 | 0.924 | 0.580 |
| 7 | 1.04 | 0.674 |
| 8 | 1.149 | 0.804 |
| 9 | 1.257 | 0.908 |

| 10 | 1.40 | 1.12 |
| 11 | 1.671 | 1.292 |
| 12 | 1.997 | 1.389 |
| 13 | 2.14 | 1.67 |
| 14 | 2.45 | 1.843 |
| 15 | 2.688 | 2.08 |
| 16 | 3.028 | 2.121 |
| 17 | 3.44 | 2.457 |
| 18 | 3.81 | 2.65 |
| 19 | 4.264 | 2.97 |
| 20 | 4.758 | 3.331 |

In Table 5, 0 explosion load refers to the displacement of the tunnel under natural excavation. It can be seen from the table that with the increase of the number of blasts, the maximum displacement of the model and the displacement of the upper collapse area of the tunnel both increase with the number of explosions. With the increase in the number of applications of explosive loads, the increment of tunnel damage displacement is also increasing.

When the fourth application of explosive load, the lowering displacement of the uppermost part of the tunnel increased by 0.064m compared to the displacement in the natural state; the maximum displacement of the tunnel increased by 0.157m. After the fourth explosion load was applied, the descent speed of the upper collapsed area of the tunnel increased. After the 6th, 12th, 16th, and 20th blasts, the descending displacement of the uppermost part of the tunnel increased by 0.596m, 1.181m, 1.913m, and 3.123m compared to the displacement in the natural state. But comparing the maximum displacement of the model with the displacement of the tunnel vault, we know that the upper part of the tunnel is not the maximum displacement of the collapse.

## 4 Analysis of tunnel failure characteristics under multiple explosion loads based on persistent homology-based machine learning

The main content of this section is the combination of SVM and tunnel failure characteristics and persistent homology. The specific process is described as follows. First, on the basis of the tunnel failure characteristics obtained through discrete element simulation in the previous section, persistent homology calculations are carried out to calculate a bar code diagram describing the tunnel failure characteristics. So far, the combination of failure characteristics and persistent homology is completed. SVM is carried out on the basis of the persistent homology calculation results. After obtaining the persistent homology characteristics of the tunnel then the explosion load is applied, machine learning is introduced to extract and recognize the topological features obtained by the persistent homology, which will reflect the surrounding rock of the tunnel. The main characteristics of the destruction process are input to the support vector machine, the previous 15 explosions are used as training samples, and the 16-20 explosions are used as test samples to predict tunnel damage. This combines SVM with tunnel failure characteristics and persistent homology.

In order to study the block displacement of the tunnel surrounding rock under the explosive load action, the surrounding rock is divided into a point cloud in the European space composed of countless blocks (it is difficult to do, generally divided into a limited number of blocks), and then get a bar code map of persistent homology topological characteristics, and then the failure characteristics of the tunnel surrounding rock are obtained based on the bar code map. Since the discrete element study , in this paper, is a two-dimensional tunnel model, only the 0-dimensional Betty number and the 1-dimensional Betty number are analyzed.

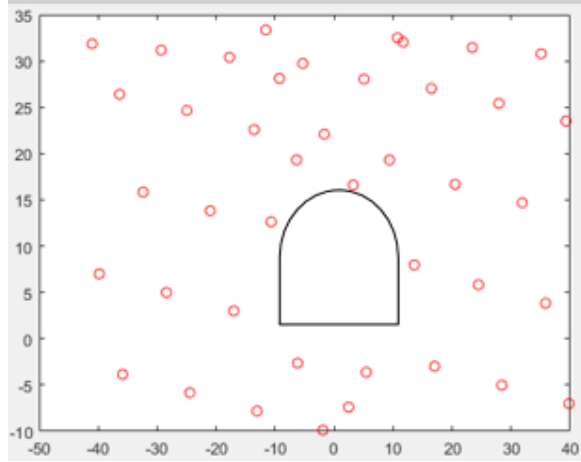

Fig.3 Scattered Point Map of Tunnel Block in Natural State

Due to the large number of explosions, all the bar code diagrams are of little significance and there are too many graphics, so only the bar code diagrams calculated under the natural excavation state and the 4th, 8th, 12th, 16th, and 20th explosions are displayed (more features have been extracted into the following graph). The nearest 42 blocks around the tunnel surrounding rock are selected as point cloud sets, and the relative areas of the selected blocks are the same. The selected areas are shown in Figure 3. Calculating the bar code graph can get the spatial homology information as follows:

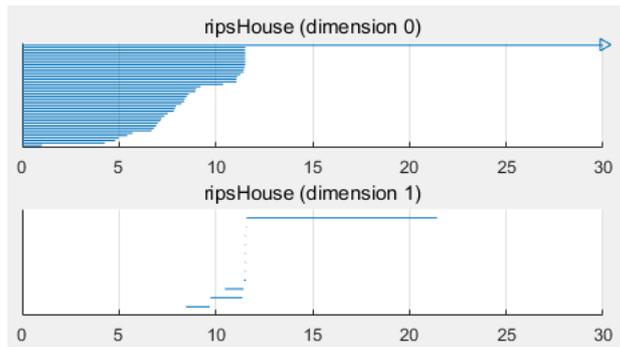

（a）natural state

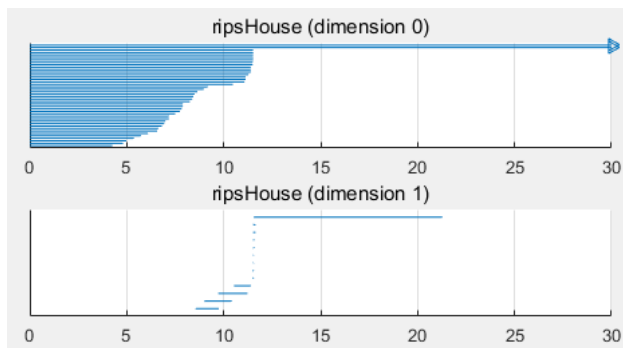

（b）The 4th explosion

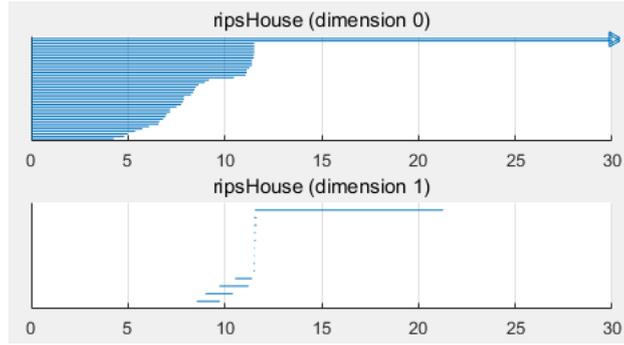

(c) The 8th explosion

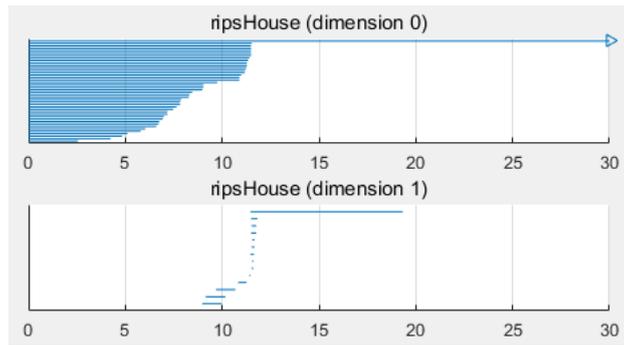

(d) The 12th explosion

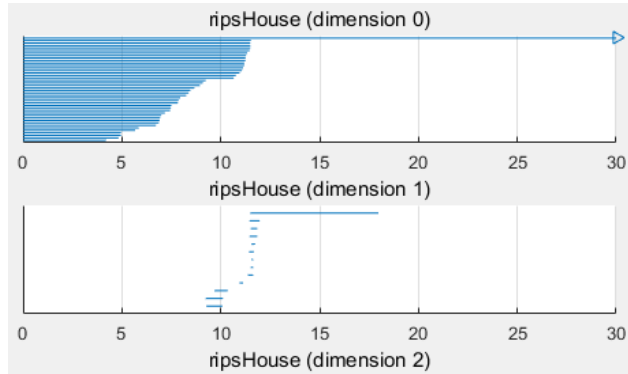

(e) The 16th explosion

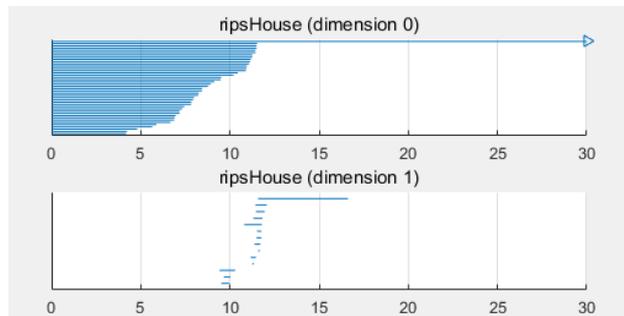

(f) The 20th explosion

Fig.4 Barcode diagram of the natural state and the 4、8、12、16 and 20 detonation tunnels

In Fig. 4, the upper layer of the bar code diagram is the 0-dimensional Betty number of bar code diagram, and the lower layer is the 1-dimensional Betty number of bar code diagram. It can be seen from the figure that the connected radius of the 0-dimensional bar code graph is [0,12], and there is almost no change from the natural state to the maximum value after each explosion. Under certain conditions of the selected block area, due to the size of the explosion load the displacement change of the edge block is relatively small. From excavation to the 20th explosion, b0=42, indicating that the number of blocks has not changed until the 20th explosion. But from the beginning of the explosion, ε began to gradually become smaller, because the explosion caused the tunnel surrounding rock to collapse, and the tunnel "holes" began to be compressed, and at this time the part of ε began to decrease.

In Fig.4, the 1-dimensional Betty number bar code diagram in the lower layer shows that with the increase number of explosions, ε always changes in an interval 8<ε<25, ε is the longest when the explosion load is applied in the natural state from time to time. A straight line begins to gradually decrease, which means that the diameter of the 1-dimensional 'holes' formed between the blocks is gradually getting smaller. This 'hole' is determined by the positional relationship of the blocks, that is, the geometry of the blocks in European space is caused by the structure. There is more than one 'hole' in this, and there are many small 'holes', which belong to the noise of the bar code graph, and we can ignore it. This section uses the persistent homology theory to analyze the dynamic change topological characteristics of the structural network of the tunnel surrounding rock structure, quantify the evolution of the macrostructure of the surrounding rock block under the action of external load, so as to better characterize the tunnel from static stability to cracking to failure. With the continuous action of the load, the tunnel is destroyed, and the length and number of the 1-dimensional Betty numbers are changing, indicating that the number of 'holes' also changes continuously with the application of explosive loads. The 0- dimensional Betty numbers also tends to become longer as a whole, indicating that the internal fissures of the rock mass structure are developing, the force chain network is deforming, and the blocks are becoming looser. It shows that there is a strong correlation between rock failure and its topological characteristics.

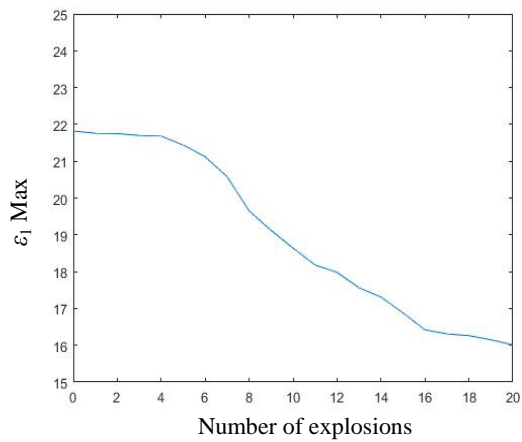

Fig.5  $\varepsilon$ maximum value changes with the number of explosions

The Fig.5 shows the maximum ε of the one-dimensional Betty number from the natural state to 1th-20th explosive loads. It can be seen from the figure that the maximum value of ε shows a decreasing trend with the increase of the explosions number , the maximum value in the natural state is 21.82, after the first explosion, the maximum value is 21.78, and it changes 0.04. The maximum value of ε after the subsequent 4, 8, 12, 16, and 20 explosions is 0.14, 2.17, 3.84, 5.4, 5.7 respectively relative to the natural state; the explosion load is applied for the 1st to 4th times, the amount of change is small. After the 5th explosion load is applied, compared with the first time, there is a change of 10 times. After the 16th explosion, the maximum change slows down. The relative change after 16-20 explosions is 0.3%.Under the action of the explosion load, the tunnel will firstly generate micro-cracks, and then the micro-cracks will continue to develop and expand, forming through cracks, until the final collapse and failure. At the same time, the blocks on both sides of the cracks are arranged tighter due to the compression effect, so that the 1-dimensional holes formed by the blocks are reduced, and the maximum value of the 1-dimensional Betty number also gradually decreases. It can be seen that the maximum value of ε in the 1D bar code graph has a strong correlation with the destruction of the tunnel surrounding rock. When the collapse zone reaches the limit of shear stress, the maximum value of ε changes drastically and accelerates. It shows that as the number of explosions increases, the displacement of the collapse increases, while the 'cavities' formed between the blocks decrease, and the collapse and destruction of the surrounding rock of the tunnel increases. After the explosion load was applied for the fourth time, the connected radius of the 1-dimensional Betty number began to change suddenly and decreased sharply toward a smaller direction. At this time, ε=21.68. When ε<21.68, continue to apply the explosive load to the tunnel, and the maximum value of ε decrease rapidly, and the damage to the surrounding rock of the tunnel increases dramatically.

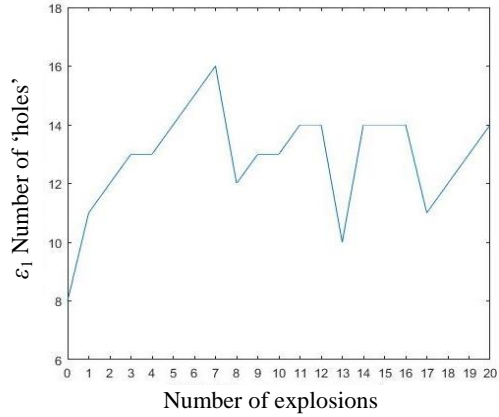

Fig.6 Total number of "holes" formed during explosion

In the natural state, the total number of "holes" in the 1D barcode map is 8, and eight "holes" are formed. After the explosion load is applied, it increases rapidly to 11, and then changes from 10 to 16, because the tunnel is destroyed under the explosion load, and the surface protruding block and internal block displacement will be formed, and the number of "holes" will change with the explosion load.

After obtaining the persistent homology characteristics of the tunnel after the applied explosion, the machine learning is introduced to extract and identify the topological features obtained by the persistent homology. The input of support vector machine mainly reflects the characteristics of the tunnel surrounding rock failure process: the interaction strength between elements and the distribution of elements, physical characteristics and geometric features, especially the geometric features, are the most intuitive that reflects the topological characteristics of surrounding rock failure process. Since this paper mainly studies the destruction features, considering the complexity of extracting all the features, only the features 2, 8, 13 and 14 in Table 1 are trained and learned, and the features 2, 8, 13 and 14 are also the quantities that can best reflect the topological features of persistent homology. In training and learning, the previous 15 blasts were used as training samples, and 16-20 times of blasts were used to test samples. In Table 6, y is the predicted value of support vector machine, j is the eigenvalue calculated and extracted by persistent homology, and W is the error between the predicted value and the calculated value.

Table 6 Support vector machine prediction results

| Explosion/time | 2 | | | 8 | | | 13 | | 14 | | |
|---|---|---|---|---|---|---|---|---|---|---|---|
| | Y | J | W | Y | J | W | Y | J | Y | J | W |
| 0 | 16.1 | | | 21.82 | | | 42 | | 8 | | |
| 1 | 16 | | | 21.76 | | | 42 | | 11 | | |
| 2 | 15.89 | | | 21.75 | | | 42 | | 12 | | |
| 3 | 15.8 | | | 21.70 | | | 42 | | 13 | | |
| 4 | 15.5 | | | 21.68 | | | 42 | | 13 | | |
| 5 | 15.4 | | | 21.44 | | | 42 | | 14 | | |
| 6 | 15.36 | | | 21.12 | | | 42 | | 15 | | |
| 7 | 15.31 | | | 20.58 | | | 42 | | 16 | | |
| 8 | 15.3 | | | 19.65 | | | 42 | | 12 | | |
| 9 | 14.56 | | | 19.12 | | | 42 | | 13 | | |
| 10 | 14.24 | | | 18.64 | | | 42 | | 13 | | |
| 11 | 13.23 | | | 18.18 | | | 42 | | 14 | | |
| 12 | 12.85 | | | 17.78 | | | 42 | | 10 | | |
| 13 | 12.64 | | | 17.56 | | | 42 | | 14 | | |
| 14 | 12.5 | | | 17.31 | | | 42 | | 14 | | |
| 15 | 12.04 | | | 16.88 | | | 42 | | 14 | | |
| 16 | 11.6 | 11.8 | 1.72% | 16.42 | 16.54 | 0.73% | 42 | 42 | 14 | 13 | 7.14% |
| 17 | 11.54 | 11.7 | 1.38% | 16.31 | 16.44 | 0.79% | 42 | 42 | 11 | 15 | 36.36% |
| 18 | 11.23 | 11.65 | 3.74% | 16.26 | 16.40 | 0.86% | 42 | 42 | 12 | 13 | 8.3% |
| 19 | 10.62 | 11.42 | 7.53% | 16.15 | 16.39 | 1.48% | 42 | 42 | 13 | 13 | 100% |
| 20 | 10.2 | 11.1 | 8.82% | 16.01 | 16.37 | 2.25% | 42 | 42 | 14 | 15 | 6.67% |

The output factors of support vector machine are the prediction results of machine learning on persistent homology in Table 6. Features 2, 8, 13 and 14 are the sum of all bar codes of 1-dimensional Betty number, the longest connecting radius length of 1-dimensional Betty number, the transverse line number of 0-dimensional Betty number, and the transverse line number of 1-dimensional Betty number, which mainly reflect the geometric characteristics and inter element action of tunnel surrounding rock under explosive load strength and distribution. It can be seen from Table 6 that the error values of feature 2, 8 and 13 are all below 10%, especially for feature 8, which reflects the failure characteristics of tunnel surrounding rock arch, the error is less than 2.5%, and the accuracy is high.

## 5 Conclusions

A new mathematical method, persistent homology-based machine learning, is introduced into the tunnel field, and combined with machine learning for intelligent research.

Based on the discrete element method (DEM), the dynamic response of the tunnel subjected to various blast loads is obtained, and the persistent homology-based machine learning method is used to study the failure characteristics. The results show that: with the increase of explosion times, the number of horizontal lines of 0-dimensional Betty number does not change, and the some 0-dimensional Betty number has a certain reduction, that is, the explosion load causes damage to the surrounding rock block and makes the connecting radius decrease to a certain extent; the maximum value of 1-dimensional Betty number is decreasing, and after the 4th explosion load is applied, that is, when the 1-dimensional

Betty number is less than the fixed value ε = 21.68. Therefore, the rapid change of maximum value of ε in 1-D bar code diagram can be used as the index and criterion to judge whether the surrounding rock of tunnel reaches the collapse boundary point.

Using the method of persistent homology-based machine learning to study the failure characteristics of tunnel surrounding rock under explosive load. The geometric structure of underground tunnel engineering can be obtained by combining with the research in the field of tunnel engineering and ground penetrating radar, the accurate and intelligent research on tunnel safety design and disaster prediction provides a new idea for tunnel research. However, the research on the stability of tunnel surrounding rock under explosion load is still in the exploratory stage, and the stability of tunnel surrounding rock under relevant conditions needs further study.

With the continuous action of the explosive load, the tunnel is destroyed, and the 0-dimensional Betty number also tends to lengthen as a whole, which indicates that the cracks in the rock structure are developing, the force chain network is deforming, and the block becomes more loose. The length and number of 1-D Betty number are changing, which indicates that the number of "holes" also changes with the application of explosion load. It shows that there is a strong correlation between rock failure and its topological characteristics.

## Acknowledgement

This work was supported by the Project Funded by Jiangxi Provincial Department of Science and Technology (No. 20192BBEL50028).